# DIRECT: Decomposing Audience Preference and Creative Effect in Visual Content Analytics

Yizhi Liu[1], Balaji Padmanabhan[2], Siva Viswanathan[2]

[1]*Fox School of Business, Temple University*

[2]*Robert H. Smith School of Business, University of Maryland, College Park*

## Abstract

Which visual choices make a post perform better? A growing literature answers this question with pooled coefficients estimated across many creators, which platforms translate into creative recommendations. We show that these coefficients blend two distinct patterns that can point in opposite directions for the same attribute. The first, audience preference, arises because creators who favor a style attract differently composed audiences, so their posts perform differently because of who is watching, not what any single post does. The second, creative effect, captures how a creator's audience responds when she departs from her usual look. Pooled estimation averages the two, and audience preference can be large enough to reverse the signal that creative direction requires. We propose DIRECT (Decomposed Identification of Response Effects via Causal Tools), a panel-based causal-inference framework that separates them, combining the Mundlak between-within decomposition with double machine learning over latent vision-language treatments that co-vary within a creator. We apply it to 232,088 sponsored Instagram beauty posts across 1,527 creators and 11 CLIP-derived visual style axes. The two carry opposite signs on 4 of 11 attributes, and on 2 of 11 the pooled coefficient itself recommends the wrong creative direction: on skin tone, it favors lighter representations while the creative effect points the other way, since a creator's audience engages more with tones darker than her baseline. On held-out creators, prescribing from the pooled coefficient forgoes 31% of the achievable engagement gain. We contribute a diagnosis of estimand mismatch in visual content analytics, a framework that recovers the decision-relevant estimand from observational panel data, and three portable diagnostics for auditing whether pooled estimates support the decisions they inform.

## Introduction

Which visual choices make a post perform better? A decade of visual content analytics research has addressed this question by extracting visual features at scale, regressing engagement on those features across many creators, and reporting per-attribute pooled coefficients (Lee et al. 2018; Li and Xie 2020; Dang et al. 2026). Pooling yields a transferable playbook of visual rules intended to apply to any creator's next post. But the pooled coefficient does not answer the question a creative decision poses. It captures which visual choices accompany better performance across creators; a creative decision requires knowing whether changing a visual choice would improve a particular creator's content. These are different questions, and we show the gap can produce not merely imprecise guidance, but directionally wrong recommendations.

To see why, consider color warmth. Creators who favor warm aesthetics may attract differently sized audiences, because warmth serves partly as a positioning signal that segments the creator market; we call this between-creator association audience preference. Within a given creator's feed, warmer-than-usual posts may elicit a different response from that creator's existing audience; we call this within-creator response creative effect. Audience preference informs creator selection; creative effect informs creative direction; and the pooled coefficient is a weighted average of the two, weighted by the data's variance structure rather than by the decision being made.

These estimates are not consumed by researchers alone. They are embedded in creative analytics systems, the scoring and recommendation tools platforms and agencies deploy to tell brands and creators what their next post should look like. Which quantity such a system estimates is a design decision, and it determines whether its recommendations are valid: one built on pooled coefficients inherits whatever mix of audience preference and creative effect the data happens to contain, and issues that mix as advice at scale, including on attributes carrying representational consequences. Designing the artifact well therefore requires specifying the estimand it should target, and giving those who already operate one a way to audit it.

We propose DIRECT (Decomposed Identification of Response Effects via Causal Tools), a causal inference framework that recovers the two separately from a single observational panel. It adapts double machine learning (Chernozhukov et al. 2018) to two features that distinguish the visual content setting from textbook panel decomposition: treatments are latent vision-language representations rather than observable scalars, and multiple visual axes co-vary within a creator, so recovering an axis-specific effect requires cross-axis nuisance modeling. Figure 1 illustrates the framework. We apply DIRECT to influencer marketing, but it fits any multi-source content panel; we write creative source for the general case and creator for our data.

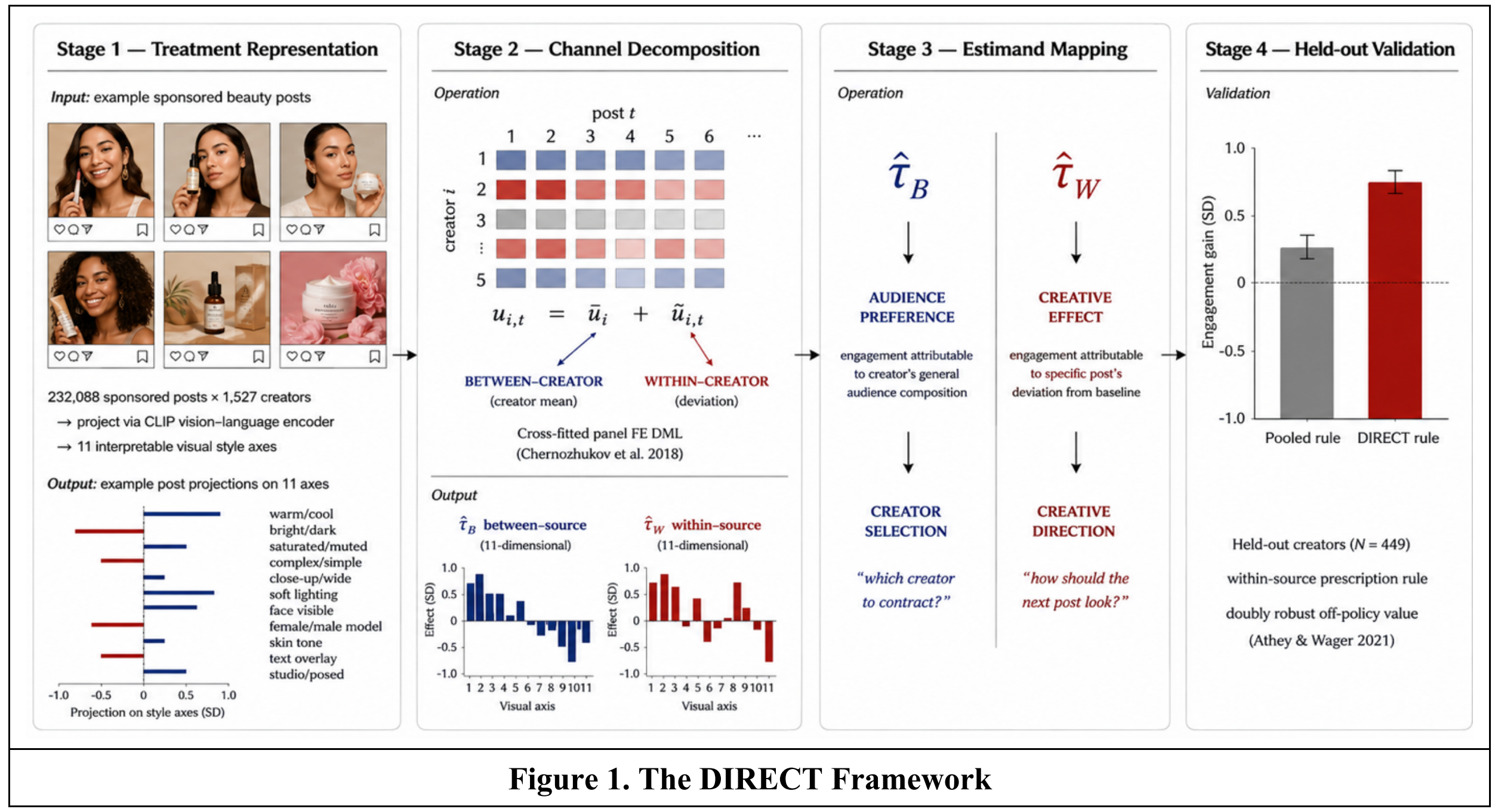

**Figure 1. The DIRECT Framework**

Applied to 232,088 sponsored Instagram beauty posts across 1,527 creators and 11 CLIP-derived visual style axes (Radford et al. 2021), warm/cool proves not to be an outlier: the two effects carry opposite signs on 4 of 11 attributes, on 2 of 11 attributes the pooled coefficient points creative direction the wrong way, and on held-out creators prescribing from the pooled coefficient forgoes 31% of the achievable within-source engagement gain under the magnitude rule.

This paper makes three contributions. First, we identify a specific form of estimand mismatch in visual content analytics: pooled coefficients blend audience preference and creative effect, and the blend can reverse the sign of creative-direction guidance. This is not bias in the usual sense, since the pooled estimate is unbiased for its own estimand, but a mismatch between the estimand reported and the decision it supports. Second, we propose a framework, built on panel-based DML, that separates the two effects from observational data. Third, we provide three portable diagnostics that let researchers with any multi-source content panel assess whether their pooled estimates support the decisions they are applied to.

## Related Work

Computational visual advertising research documents how visual features drive engagement on social media (Lee et al. 2018; Li and Xie 2020; Dang et al. 2026), sharing one empirical strategy: extract visual features at scale, then estimate their relationship with engagement using pooled specifications with covariate controls. That strategy describes cross-creator patterns consistently, but it does not separate the two economic effects underlying engagement, a separation that matters because the source of content itself shapes consumer response (Goh et al. 2013; Cao and Belo 2024).

Our methodological core is panel-based double machine learning (Chernozhukov et al. 2018), applied in marketing by Ellickson et al. (2023) and, for policy learning on observational data, by Athey and Wager (2021). We adopt this decision-theoretic stance but at a different level: not the choice of whom to treat, but which estimand should serve as the basis for treatment design. Our estimator combines DML with the Mundlak (1978) decomposition, which supplies the algebraic identity between pooled, between, and within estimands but does not, on its own, yield axis-specific marginal effects when treatments are latent vision-language projections that co-vary across style dimensions. The distinction is not a formality: the benchmark reported below shows that within-creator fixed effects without cross-axis orthogonalization inflates magnitudes by 40 to 100% on most axes, by roughly 1,300% on one, and reverses the sign on 2 of our 11 axes.

## The DIRECT Framework

### *Two Decisions, Two Estimands*

Content production decisions occur at two levels: selecting a creative source, and directing its output. Once a source is contracted, its identity, audience, and baseline aesthetic are fixed, and the remaining lever is the creative brief: warmer tones for the autumn line, feature darker shades, reduce text overlay. Formally, the advertiser chooses a visual style u to maximize E[Y | source i, brief b, style u]; since i and b are fixed at decision time, the first-order condition on attribute k is the within-source partial effect of $u_k$ on Y.

**Definition 1 (Direction Estimand).** The direction-relevant causal estimand for visual style attribute k is the within-source partial effect of $u_k$ on engagement Y, denoted $\tau_W$. This is the partial effect a creative brief operationalizes; its identification as a causal effect relies on the conditional-exogeneity assumption stated below.

Variation in any style attribute has two origins: between-source differences in baseline style, which correlate with audience composition and other source-level factors, and within-source differences across posts by the same source. The two can move an attribute in opposite directions.

### *The Pooled Coefficient and Its Decomposition*

The standard approach regresses outcomes on visual attributes across the full sample and reports a single pooled coefficient per attribute. Under panel structure, this coefficient is a variance-weighted average of the between-source $\tau_B$ and the within-source $\tau_W$ (Mundlak 1978):

$$\tau_{\text{pool}} = w_B \cdot \tau_B + w_W \cdot \tau_W$$

The weights reflect each attribute's between- and within-source variance shares; they are properties of the data, not choices of the analyst. The pooled coefficient answers a descriptive question correctly, but creative direction requires $\tau_W$ alone. When pooled estimates are read as direction guidance, as is routine in trade press, industry tools, and implications sections, the rule implicitly treats $\tau_B$ as if it captured creative effect.

The discrepancy is $\tau_{\text{pool}} - \tau_W = w_B \cdot (\tau_B - \tau_W)$, zero only when the data has no between-source variation ($w_B = 0$) or the two channels coincide ($\tau_B = \tau_W$); both are typically violated in observational panels. Because our pooled estimate comes from a separate full-sample DML rather than being reconstructed from the two channels, this identity need not hold exactly in finite samples. Its implication is what matters: where $\tau_B$ and $\tau_W$ disagree in sign, the pooled rule can point opposite to the $\tau_W$ rule, and where they disagree in magnitude, it is directionally correct but mis-sized.

### *The Four Stages of DIRECT*

DIRECT proceeds in four stages, summarized in Figure 1. Stage 1, treatment representation, projects images into K interpretable visual style axes via a pre-trained vision-language encoder; axes must pass face validity and stability across encoder versions. Stage 2, channel decomposition, applies the Mundlak device to split each attribute into a source-level mean and a post-level deviation from it, and panel-based DML then jointly identifies $\tau_B$ and $\tau_W$ per axis, with nuisance functions absorbing post-level controls, source-level covariates, and cross-axis confounding from the remaining K−1 axes. Stage 3, estimand selection, maps $\hat{\tau}_B$ to creator selection and $\hat{\tau}_W$ to creative direction, the estimand a creative brief operationalizes. Stage 4, decision-relevant validation, asks whether following the $\hat{\tau}_W$ rule instead of the $\hat{\tau}_{pool}$ rule recovers more engagement on held-out creators, reporting the conflation cost in percentage terms, with off-policy doubly-robust value estimation (Athey and Wager 2021) as an independent check.

### *Diagnostics and Conditions of Applicability*

DIRECT produces three portable diagnostics. D1, the channel-disagreement count, is the number of axes on which $\hat{\tau}_B$ and $\hat{\tau}_W$ disagree in sign, both above a magnitude floor; D1′ counts axes on which $\hat{\tau}_{pool}$ itself disagrees in sign with $\hat{\tau}_W$, so pooled-based guidance points the wrong way. The two counts need not nest: because the pooled estimate comes from a separate full-sample fit rather than from the two channels, an axis can fall in D1′ without falling in D1. D2, the magnitude divergence ratio $|\hat{\tau}_B - \hat{\tau}_W| / \max(|\hat{\tau}_{pool}|, \varepsilon)$, flags disagreement exceeding what pooled estimation reports even when signs agree. D3, the direction-prescription cost, is the percentage effectiveness loss from prescribing on pooled rather than within-source estimates.

The framework applies wherever an identifiable source contributes multiple units, the intervention varies the act rather than the actor, the decision-relevant attributes are extractable from the unit via a pre-trained encoder, and the outcome is observed per unit. $\hat{\tau}_W$ approximates the within-source causal effect only under a further assumption: conditional on source fixed effects, post-level controls, and cross-axis nuisance modeling, residual within-source variation in $u_k$ is approximately exogenous to residual variation in Y. This is not mechanical: within-source variation can still reflect post-level selection on campaign objectives, product types, or sponsorship terms. Orthogonalization against high-dimensional controls and the held-out validation of Stage 4 are what we offer against it.

## Empirical Implementation

### *Data, Visual Representation, and Estimation*

Our analysis uses 232,088 sponsored Instagram beauty-category posts contributed by 1,527 creator accounts over a 24-month window spanning 2024–2025. The beauty category exhibits substantial within-creator stylistic variation. Creators contribute a mean of 152 posts (median 162), and the panel spans the full size distribution, from 524 micro accounts (1k–10k followers) to 71 mega (above 1M), median 18,325 followers. Engagement is highly skewed (median 572 likes, mean 5,999, SD 30,819), so the outcome is log(1 + likes).

We project each post image into 11 face-validated visual style axes using CLIP ViT-L/14 (Radford et al. 2021), drawn from a literature-driven pool spanning color, lighting, composition, framing, and demographic representation (listed in Table 1). Each axis is the difference in cosine similarity between the image embedding and a pair of natural-language anchor prompts at the axis extremes. Face validity was checked by inspecting the 50 highest- and 50 lowest-scoring posts per axis; scores are stable across encoders (ViT-L/14 to ViT-B/16 rank correlations above 0.92).

For each axis k we estimate the two-level partial linear model

$$Y_{i,t} = \tau_{\mathrm{B}}(k)\, \bar{u}_i(k) + \tau_{\mathrm{W}}(k)\, \tilde{u}_{i,t}(k) + g_k(X_{i,t}, Z_i, u_{i,t}(-k)) + \varepsilon_{i,t},$$

where $\bar{u}_i(k)$ is source i's mean of attribute k, $\tilde{u}_{i,t}(k)$ is the within-source deviation from it, and $g_k$ is an unknown nuisance function absorbing post-level controls $X_{i,t}$ (posting time, aspect ratio, face count and confidence), source-level covariates $Z_i$ (follower and post counts, verified status, tier dummies), and the remaining $K-1 = 10$ axes. Each channel is recovered from its own Neyman-orthogonal score, so first-stage regularization does not contaminate inference on ($\tau_B$, $\tau_W$). Nuisance functions are LightGBM regressors cross-fitted with 5-fold group K-fold keyed on source ID, standard errors are cluster-robust by source, and $\hat{\tau}_{\mathrm{pool}}$ comes from a separate full-sample DML.

### ***Held-Out Validation Protocol***

Validation uses a creator-disjoint train/hold split, so rules estimated on train creators (1,068 creators, 161,382 posts) are applied to hold-out posts of hold-out creators. The hold-out data (459 creators, 70,706 posts) splits within-creator into a signal window estimating creator means and an outcome window used as the prediction target, giving a validation panel of 35,336 posts from 449 creators. A prescription rule maps each estimand to an attribute-aggregated direction score, weighting each within-source deviation by the sign of the estimate (sign rule) or by its value (magnitude rule). Regressing the outcome deviation on each score, clustered by source, gives slopes $\beta_r$, and the pooled rule's relative loss is $100 \cdot (\exp(\beta_W) - \exp(\beta_{pool})) / (\exp(\beta_W) - 1)$.

## Results

### *Attribute-Level Estimates*

Table 1 reports the decomposition for all 11 axes: the between-source $\hat{\tau}_B$, the within-source $\hat{\tau}_W$, the pooled estimate $\hat{\tau}_{pool}$, and the implied sign agreement. Two patterns emerge. First, $\hat{\tau}_B$ and $\hat{\tau}_W$ reverse signs on 4 of 11 axes, namely warm/cool, female/male model, skin tone, and selfie/posed, so the pooled coefficient blends opposing economic signals there. Second, the within-source channel disagrees in sign with the pooled estimate on 2 of 11 axes: on skin tone the pooled estimate is significant and points the opposite way, and on close-up/wide it is null while within-source identification points clearly the other way. In both cases, prescriptions read off pooled estimates fail to recover the within-source direction; on the 9 axes where the two agree in sign, magnitudes still differ substantially, mis-calibrating the size of the adjustment even where its direction is right.

| Visual Style Axis | $\tau_B$ | $\tau_W$ | $\tau_{pool}$ | $\tau_B$ vs. $\tau_W$ |
|---|---|---|---|---|
| 1. Warm vs. cool | −3.90*** | +0.18* | +0.10 (n.s.) | Disagree |
| 2. Bright vs. dark | +3.95*** | +0.84*** | +2.81*** | Agree |
| 3. Saturated vs. muted | +3.20*** | +0.65*** | +0.20 (n.s.) | Agree |

| Visual Style Axis | $\tau_B$ | $\tau_W$ | $\tau_{pool}$ | $\tau_B$ vs. $\tau_W$ |
|---|---|---|---|---|
| 4. Complex vs. simple | −3.05*** | −0.07* | −1.04* | Agree |
| 5. Close-up vs. wide | −2.71*** | −0.47*** | +0.09 (n.s.)‡ | Agree |
| 6. Soft vs. hard lighting | −8.50*** | −0.21* | −2.91*** | Agree |
| 7. Face visible vs. product only | +2.81*** | +0.14* | +0.08 (n.s.) | Agree |
| 8. Female vs. male model | −0.84* | +0.99*** | +0.24 (n.s.) | Disagree |
| 9. Light vs. dark skin tone | +2.13*** | −0.64*** | +0.95*‡ | Disagree |
| 10. Text overlay presence | −9.04*** | −2.68*** | −5.39*** | Agree |
| 11. Selfie vs. posed | +2.24*** | −0.33* | −0.14 (n.s.) | Disagree |

Notes: $\tau_B$ from between-source DML (N = 1,527 creators); $\tau_W$ from within-source DML (N = 232,088 source-demeaned posts); $\tau_{pool}$ from full-sample pooled DML. *p < 0.05; **p < 0.01; ***p < 0.001, cluster-robust by source. ‡ marks axes where $\tau_{pool}$ reverses sign against $\tau_W$. Coefficients are per CLIP-axis unit; percentage effects quoted in the text convert these using the corresponding standard deviation of the axis.

**Table 1. Attribute-Level Channel Decomposition**

**Text overlay and warm/cool are the milder failures.** On text overlay both channels agree on the negative sign but not on size: $\hat{\tau}_{pool} = -5.39$ is about a 15% reduction per standard deviation, $\hat{\tau}_W = -2.68$ about 7% per within-creator standard deviation, and most of the apparent penalty is the between-source pattern $\hat{\tau}_B = -9.04$, since creators who use more text overlay have systematically smaller audiences. A team setting reduction targets from the pooled coefficient therefore overestimates the available gain, and warm/cool is the mirror image: $\hat{\tau}_{pool} = +0.10$ reads as warmth being irrelevant, yet it is approximately $\hat{\tau}_B = -3.90$ cancelling a positive $\hat{\tau}_W = +0.18$.

**Skin tone is a sign reversal with representational implications.** Pooled estimation reports $\hat{\tau}_{pool} = +0.95$ ($p < 0.05$), about 5% higher engagement per standard deviation toward lighter skin tone, which reads as a directive to feature lighter tones. Within-source identification reverses the sign: $\hat{\tau}_W = -0.64$ ($p < 0.001$), so within a creator’s content stream, posts featuring darker skin tones receive higher engagement than that creator’s average (about 1.8% per within-creator standard deviation), while $\hat{\tau}_B = +2.13$ dominates the pooled blend. The channels measure different phenomena. The between-source signal reflects an audience-preference pattern in our panel, in which creators with lighter baseline skin tones tend to have larger followings; creative direction cannot act on it, since it cannot change a contracted creator's skin tone or audience composition. The within-source signal reflects how a creator's audience responds when that creator features darker tones in a given post; that effect is consistent with, but does not by itself prove, a causal interpretation. What is unambiguous is that pooled-based guidance for skin tone would systematically over-weight lighter representations, on a basis that is not creative effect but audience preference. Close-up versus wide framing follows the same logic on a smaller scale ($\hat{\tau}_{pool} = +0.09$, n.s.; $\hat{\tau}_W = -0.47$, $p < 0.001$).

### *Held-Out Validation of Creative-Direction Rules*

Table 2 reports held-out validation results. Pooled rules deliver +6.94 to +7.94% engagement gain per standard deviation, while DIRECT rules deliver +8.48 to +11.47%. Read as the cost of using the pooled rule instead of DIRECT, the pooled rule forgoes 30.7% of the achievable gain under the magnitude rule and 18.2% under the sign rule. An off-policy doubly-robust value estimator corroborates the ranking on a metric that is not on the same scale as the slope: $\hat{V}_{DR}(\pi_W)$ = +3.67% engagement versus random allocation, against $\hat{V}_{DR}(\pi_{pool})$ = +2.99%.

| Prescription Rule | Engagement Gain (% per SD) | t-stat |
|---|---|---|
| Pooled-CATE (sign rule) | +6.94 | +11.63 |
| Pooled-CATE (magnitude rule) | +7.94 | +11.87 |
| DIRECT $\hat{\tau}_W$ (sign rule) | +8.48 | +11.32 |
| DIRECT $\hat{\tau}_W$ (magnitude rule) | +11.47 | +15.73 |
| Notes: Held-out validation on 35,336 posts from 449 hold-out creators. Engagement gain is percent per standard deviation of the attribute-aggregated direction score; standard errors cluster-robust by source. | | |

**Table 2. Held-Out Validation of Creative-Direction Rules**

### *Estimator Benchmark and Robustness*

The between-within decomposition is itself standard; what this setting adds is the cross-axis modeling around it. A benchmark shows how much it matters: within-creator fixed-effects ordinary-least-squares, which demeans by source exactly as DIRECT does but omits cross-axis nuisance modeling, agrees in sign on 9 of 11 axes and disagrees on two, soft lighting and selfie/posed, where it returns the opposite creative direction. Where signs agree its magnitudes exceed DIRECT's by 40 to 100%, and by roughly 1,300% on complex-versus-simple. Because visual axes co-vary strongly within a creator, demeaning alone attributes joint variation to whichever axis is in focus; orthogonalizing against the others recovers the partial effect appropriate to a directive that adjusts one attribute holding the rest fixed.

Latent treatments carry a second consequence. Because the CLIP axes measure visual style with error, source-demeaning lowers the signal-to-noise ratio and attenuates within-source estimates toward zero, one reason $|\hat{\tau}_W|$ is smaller than $|\hat{\tau}_B|$ on 10 of 11 axes. Attenuation cannot reverse signs, so the sign reversals our claims rest on are unaffected, while the magnitude comparisons and D2 are best read as upper bounds.

Estimates preserve $\tau_W$ signs on 11 of 11 axes across fold choices $K \in \{3, 5, 10\}$ and LightGBM n_estimators $\in \{100, 300, 500\}$. Disjoint 50% creator subsamples correlate at $\rho(\tau_W) = 0.996$ and $\rho(\tau_B) = 0.872$; a placebo test permuting the outcome rejects the null on no axis at the Bonferroni-corrected 5% level; and skin tone and close-up/wide remain negative across all five follower tiers (skin tone −0.28 to −0.81), with the text-overlay penalty markedly larger for large accounts than for micro accounts (−3.73 against −2.07). Under the omitted-variable-bias framework of Cinelli and Hazlett (2020), an unobserved within-creator confounder would need to explain more residual variation in both the attribute and engagement than the strongest observed control, by roughly 8 times for skin tone, 15 times for female/male model, and 5 times for close-up/wide framing. These are benchmarks relative to observed confounding, not absolute guarantees: no single visual attribute explains more than a fraction of a percent of within-creator engagement variation. Applied to our panel, the diagnostics are D1 = 4 of 11 axes, D1′ = 2 of 11, D2 > 1.0 on 11 of 11, and D3 = 30.7%.

## Discussion and Implications

For practice, a creative analytics system should report the two channels separately, $\hat{\tau}_W$ as the creative-effect signal and $\hat{\tau}_B$ as audience-preference context, rather than the single leaderboard that serves both decisions at once. The diagnostics give a fast screen: on axes outside D1′ (9 of 11 here) pooled prescription is directionally correct though possibly mis-sized; on axes within it, decomposition is required.

For research, pooled estimates remain valid descriptive statistics; what the framework adds is a separation of the two effects they average together. Papers reporting attribute-engagement coefficients should state whether they are meant descriptively or prescriptively, and report panel-based decomposition where the latter is intended. Our 4-of-11 reversal rate and 30.7% conflation cost are specific to this panel, but they suggest the gap matters for findings already built into platform tooling. The skin-tone case connects this work to algorithmic bias in commercial systems (Buolamwini and Gebru 2018): DIRECT does not eliminate fairness concerns in visual advertising, but it makes visible a distinction pooled estimation hides, that positive pooled coefficients on lighter representations may reflect audience-recruitment history rather than creative effectiveness.

Two limitations bound the analysis: residual within-source confounding cannot be fully ruled out, and the data cover one category, platform, and time window. A within-creator natural experiment and cross-category replication would address both.